\documentclass[useAMS,usenatbib]{mn2e}
\usepackage{amsmath,amssymb}
\usepackage{graphicx,subfigure}
\usepackage{multirow}
\pdfoutput=1
\usepackage{hyperref}
\usepackage{epstopdf}

\newcommand{\bb}{\begin{bmatrix}}
\newcommand{\eb}{\end{bmatrix}}

\begin{document}
\pagerange{\pageref{firstpage}--\pageref{lastpage}} \pubyear{2016}

\title[Detection and localization of continuous GWs with PTAs]{Detection and localization of continuous gravitational waves with pulsar timing arrays: the role of pulsar terms}
\author[X.-J. Zhu et al.]
{X.-J. Zhu$^{1}$\thanks{E-mail: zhuxingjiang@gmail.com},
L. Wen$^{1}$,
J. Xiong$^{1,2}$,
Y. Xu$^{1,3}$,
Y. Wang$^{4}$,
S. D. Mohanty$^{5}$,
G. Hobbs$^6$,
\newauthor
R. N. Manchester$^6$
\\
$^1$ School of Physics, University of Western Australia, Crawley WA 6009, Australia\\
$^{2}$ Department of Modern Physics, University of Science and Technology of China, Hefei 230026, China\\
$^{3}$ Department of Astronomy, University of Science and Technology of China, Hefei 230026, China\\
$^{4}$ School of Physics, Huazhong University of Science and Technology, 1037 Luoyu Road, Wuhan, Hubei 430074, China\\
$^{5}$ Department of Physics, The University of Texas Rio Grande Valley, 1 West University Boulevard, Brownsville, TX 78520, USA\\
$^{6}$ CSIRO Astronomy and Space Science, PO Box 76, Epping NSW 1710, Australia\\
}
\maketitle \label{firstpage}

\begin{abstract}
A pulsar timing array is a Galactic-scale detector of nanohertz gravitational waves (GWs). Its target signals contain two components: the `Earth term' and the `pulsar term' corresponding to GWs incident on the Earth and pulsar respectively. In this work we present a Frequentist method for the detection and localization of continuous waves that takes into account the pulsar term and is significantly faster than existing methods. We investigate the role of pulsar terms by comparing a full-signal search with an Earth-term-only search for non-evolving black hole binaries. By applying the method to synthetic data sets, we find that (i) a full-signal search can slightly improve the detection probability (by about five percent); (ii) sky localization is biased if only Earth terms are searched for and the inclusion of pulsar terms is critical to remove such a bias; (iii) in the case of strong detections (with signal-to-noise ratio $\gtrsim$ 30), it may be possible to improve pulsar distance estimation through GW measurements.
\end{abstract}


\section{Introduction}
The recent detection of gravitational waves (GWs) from a stellar-mass binary black hole merger by Advanced LIGO opened up a brand new window for observational astrophysics \citep{GW150914}. With steadily improved sensitivity for Advanced LIGO \citep{LVC13localiz} and the addition of other advanced ground-based laser interferometers such as Advanced Virgo \citep{aVirgo} and KAGRA \citep{KAGRA}, a large number of detections are expected in the audio frequency band (10--1000 Hz) in the coming years. However, to have a comprehensive view of the gravitational-wave Universe, observations on other parts of the GW spectrum are critical. A pulsar timing array \citep[PTA;][]{Foster_Backer90}, which involves long-term timing observations of a spatial array of millisecond pulsars, provides an unique way to detect GWs in the nanohertz band (1--100 nHz). Such experiments are being carried out by three international collaborations -- the Parkes PTA \citep[PPTA;][]{PPTA2013,PPTA13CQG} in Australia, the European PTA \citep[EPTA;][]{EPTA}, and the North American Nanohertz Observatory for Gravitational Waves \citep[NANOGrav;][]{NANOGrav}. These collaborations have also combined to form the International PTA \citep[IPTA;][]{IPTA,IPTAdick13}, for which the first data release was published very recently \citep{IPTA16}.

Potential sources of GWs that are detectable by PTAs include inspiralling supermassive black hole binaries (SMBHBs) in the centres of galaxies \citep{Rajago95,Jaffe_Backer03,Wyithe_Loeb03,Sesana13GWB,Ravi14GWB}, cosmic strings \citep{Vilenkin81}, and primordial gravitational fluctuations amplified by an inflationary phase \citep{Grishchuk05}. Previous efforts have mostly gone into searching for a stochastic background, either from cosmological sources or formed by the combined emission of SMBHBs distributed throughout the Universe. Over the past decade, the strain sensitivity to such background signals has been increased by an order of magnitude, leading to very stringent constraints on early-Universe physics and characteristics of the SMBHB population \citep{Jenet2006,epta15GWB,NANO15gwb,PPTA15Sci}.

Over the past few years, interest has grown substantially in detecting and studying individual GW sources with PTAs \citep{Sesana2010,Finn10ApJ,KJLee2011,Babak2012,Ellis2012,Zhu15Single,YanWang14,YanWang15,ApAc16}. In the case of SMBHBs, the closest and most massive binaries are naturally expected to produce GWs that exceed the level of the background and can be individually detectable. Detailed simulations have shown that the background may be dominated by a small number of single sources \citep{Sesana2009,Ravi2012}, and that there is a sizable chance of detecting individual binaries given sufficiently improved sensitivity \citep{Ravi14Single,Rosado15}. \citet{Yardley2010} produced the first sensitivity curve and set upper limits for single-source GWs. More recently, three PTAs performed searches for GWs from individual SMBHBs in circular orbits and obtained stringent limits on the GW strain amplitude and on the local binary coalescence rate \citep{NANOcw14,PPTAcw14,EPTAcw15}.

GWs passing across the pulsar-Earth baseline perturb the local spacetime along the path of radio wave propagation, leading to fluctuations in the arrival times of radio pulses from millisecond pulsars. Because the gravitational wavelength is much shorter than the typical pulsar distance (thousands of light years), GW signals buried in PTA data consist of two terms: the ``pulsar term" and the ``Earth term", corresponding to GWs incident on the pulsar and the Earth respectively \citep{Detweiler1979,JenetWen04}. The Earth term is coherent for all pulsars being monitored in the array, while the phase of the pulsar term depends on pulsar distance which is poorly known for most pulsars. A majority of previous work on searches for GWs from individual SMBHBs treated pulsar terms as a source of self-noise and thus ignored them in the search and parameter estimation algorithms \citep[e.g.,][]{Sesana2010,Babak2012,Petiteau13,Zhu15Single,ApAc16,Taylor16ecc}. However, a coherent inclusion of pulsar terms has been suggested to be critical for improving detection probability and sky localization \citep{KJLee2011,ULPen12PRD}. Additionally, the detection of pulsar terms may help probe the binary evolutionary histories \citep{Chiara12PRL} and improve the pulsar distance measurements \citep{KJLee2011}.

In this work, we extend our Earth-term-based data analysis method developed in \citet{Zhu15Single} to include a coherent search for pulsar terms from individual SMBHBs in circular orbits and during their early inspiral stages. The inclusion of pulsar terms increases the search dimension by the number of pulsars in the array and is computationally limiting for Bayesian techniques \citep[e.g.,][]{EllisBayesian,Taylor14}. Here, we present a fast Frequentist method that quickly returns reliable results of detection and unbiased sky localization, and more importantly can be applied to an array with tens of pulsars without causing significant computational overload. Furthermore, using a simulated IPTA data set we quantify (1) how the inclusion of pulsar terms can improve the detection probability and (2) how the sky localization is biased for an Earth-term-only search. A Frequentist approach that takes pulsar terms into account is presented in \citet[][hereafter WMJ15]{YanWang15} but it estimates them in a significantly different manner: while our method estimates the pulsar terms numerically, the method of WMJ15 estimates them analytically. Consequently, noise in the data is processed very differently in the two methods, leading to differences in pulsar term estimation errors.

The organization of this paper is as follows. In Section \ref{sigmodel} we review the response of a PTA to single-source GWs and describe the signal model used in this work. In Section \ref{SVDmethod} we describe our detection method and the technique used to conduct high-dimensional searches. In Section \ref{sec-results} we first show its applications in detection and parameter estimation for different signal-to-noise ratios using synthetic data sets. We then compare the new method to Earth-term-only approaches and quantify improvements in terms of detection and sky localization. We also compare the computation speed with other published methods. Finally Section \ref{Conclu} contains our conclusions.

\section{The signal}
\label{sigmodel}
A PTA data set is composed of time of arrival (ToA) measurements for pulses of those millisecond pulsars included in the timing array. Such observations are made over many years with a typical sampling interval of weeks, implying a PTA band of $\sim$1--100 nHz for GW detection. The measured ToAs are fitted to a timing model that describes the pulsar's rotational behavior such as its spin period and spin-down rate, the astrometry of the pulsar and other effects including the presence of a binary companion and the dispersion of radio waves due to ionized interstellar medium. Timing residuals are then obtained by subtracting measured ToAs with the model-predicted ToAs. Major contributions to timing residuals are various noise processes and GWs. A simple measure of quality for pulsar timing data is the root mean square (rms) of the timing residuals, which is $\lesssim$ 100 ns for a few pulsars and $\lesssim$ 1 $\mu$s for most pulsars currently timed by the three PTAs.

Consider a single GW source coming from a direction $\hat{\Omega}$, its induced pulsar timing residuals measured at time $t$ on the Earth can be written as:
\begin{equation}
\label{TR1}
s(t,\hat{\Omega}) = F^{+}(\hat{\Omega})\Delta A_{+}(t) + F^{\times}(\hat{\Omega}) \Delta A_{\times}(t),
\end{equation}
where $F^{+}(\hat{\Omega})$ and $F^{\times}(\hat{\Omega})$ are the antenna pattern functions as given by \citep{Wahlq87}:
\begin{eqnarray}
F^{+}(\hat{\Omega}) &=& \frac{1}{4(1-\cos\theta)}\{(1+\sin^2 \delta)\cos^2 \delta_p \cos[2(\alpha-\alpha_p)]\nonumber\\&&\hspace{-14mm} - \sin2\delta \sin2\delta_p\cos(\alpha-\alpha_p) + \cos^2 \delta (2-3\cos^2 \delta_p)\}
\label{Fp}
\end{eqnarray}
\begin{eqnarray}
F^{\times}(\hat{\Omega}) &=& \frac{1}{2(1-\cos\theta)}\{\cos \delta \sin 2\delta_p \sin(\alpha-\alpha_p)\nonumber\\&& - \sin \delta \cos^2 \delta_p \sin[2(\alpha-\alpha_p)]\} ,
\label{Fc}
\end{eqnarray}
where $\alpha$ ($\alpha_p$) and $\delta$ ($\delta_p$) are the right ascension and declination of the GW source (pulsar) respectively, and $\theta$ is the opening angle between the GW source and pulsar with respect to the observer
\begin{equation}
\label{costheta}
\cos\theta = \cos\delta \cos\delta_p \cos(\alpha-\alpha_p)+\sin\delta \sin\delta_p \, .
\end{equation}

In equation (\ref{TR1}), we define
\begin{equation}
\label{TR2}
\Delta A_{\{+,\times\}}(t) = A_{\{+,\times\}}(t)-A_{\{+,\times\}}(t_p),
\end{equation}
where $t_p$ is the time at which the GW passes the pulsar
\begin{equation}
\label{TpTe}
t_p = t-d_p(1-\cos\theta)/c,
\end{equation}
where $d_p$ is the pulsar distance. Here $A_{\{+,\times\}}(t)$ and $A_{\{+,\times\}}(t_p)$ contribute to the Earth term and pulsar term respectively, for which the specific functional forms depend on the type of sources being searched for. For SMBHBs in circular orbits, they are given by
\begin{eqnarray}
A_{+}(t) &=& \frac{h_0}{2\pi f(t)} \{(1+\cos ^{2}\iota) \cos2\psi \sin[\phi(t)+\phi_0]\nonumber\\&& +2\cos\iota \sin2\psi \cos[\phi(t)+\phi_0]\}
\label{Ap}
\end{eqnarray}
\begin{eqnarray}
A_{\times}(t) &=& \frac{h_0}{2\pi f(t)} \{(1+\cos ^{2}\iota) \sin2\psi \sin[\phi(t)+\phi_0]\nonumber\\&& -2\cos\iota \cos2\psi \cos[\phi(t)+\phi_0]\}.
\label{Ac}
\end{eqnarray}
Here $\iota$ is the inclination angle of the binary orbital plane with respect to the line of sight, $\psi$ is the GW polarization angle, $\phi_0$ is a phase constant, and $h_0$ is the intrinsic GW strain amplitude defined as
\begin{equation}
h_{0}=2\frac{(G M_c)^{5/3}}{c^{4}}\frac{(\pi f)^{2/3}}{d_{L}}
\label{h0},
\end{equation}
where $d_L$ is the luminosity distance to the source, and $M_c$ is the binary chirp mass defined as $M_c^{5/3} = m_1 m_2(m_1+m_2)^{-1/3}$ with $m_1$ and $m_2$ being the binary component masses. It should be noted that it is the \emph{redshifted} chirp mass $M_c^{z}=M_{c}(1+z)$ that is directly measurable in GW detection; likewise, the rest-frame frequency $f_r$ is related to the observed frequency by $f_{r}=f(1+z)$. \citet{Pablo16} recently showed that PTAs are equally sensitive to SMBHBs at high redshift ($z\gtrsim 2.6$) and those at the local Universe given that both are observed at the same frequency. We will focus on nearby systems in this work and leave the search for high-$z$ binaries to a future study.

In the quadrupole approximation, the GW phase and frequency that appear in equations (\ref{Ap}-\ref{Ac}) are given by \citep{Thorne87}:
\begin{equation}
\label{GW-Phase}
\phi(t) = \frac{1}{16}  \left(\frac{G M_c}{c^{3}}\right)^{-5/3}\left\{(\pi f_{0})^{-5/3}-[\pi f(t)]^{-5/3}\right\},
\end{equation}
\begin{equation}
\label{GW-freq}
f(t) = \left[f_{0}^{-8/3}-\frac{256}{5}\pi^{8/3}\left(\frac{G M_c}{c^{3}}\right)^{5/3}t\right]^{-3/8},
\end{equation}
where $f_0$ is the GW frequency at the time of our first observation. From equation (\ref{GW-freq}), the frequency evolution over a short time scale is given by
\begin{equation}
\label{dfevo}
\Delta f \simeq 3.94\,{\rm{nHz}} \left(\frac{M_{c}}{10^{9}\, M_{\odot}}\right)^{5/3}\left(\frac{f_0}{10^{-7}\,{\rm{Hz}}}\right)^{11/3}\left(\frac{T_{\rm{obs}}}{10\, {\rm{yr}}}\right)\, .
\end{equation}
For a typical PTA data span $T_{\rm{obs}}\sim$ 10 yr, this is smaller than the frequency resolution of PTA observations except for the high-mass ($>10^{9}\, M_{\odot}$) and high-frequency ($>100$ nHz) regime. Folding in the fact that binary systems spend most of their lifetime at low frequencies (${\rm{d}}f/{\rm{d}}t \sim f^{11/3}$), monochromatic binaries indeed represent an overwhelming majority of detectable sources.

In this work, we focus on non-evolving binaries for which the frequency evolution within the pulsar-Earth baseline -- $d_p(1-\cos\theta)/c$ -- is negligible ($< 1/T_{\rm{obs}}$). It has been shown that such binaries represent about $50\%-78\%$ of detectable sources for simulated populations of SMBHBs for a PTA with pulsars $\lesssim 1.5$ kpc away \citep{Rosado15}. Under these assumptions, the full signal is the sum of two sinusoids of different phases as given by
\begin{eqnarray}
s(t,\hat{\Omega}) &=& \frac{h_0}{\pi f}\sin(\Delta\Phi/2) \{[F^{+}(\hat{\Omega})\cos(2\psi)+F^{\times}(\hat{\Omega})\sin(2\psi)] \nonumber\\&& (1+\cos ^{2}\iota)\cos(2\pi f t+\phi_0-\Delta\Phi/2) \nonumber\\&&  +2\cos\iota [F^{\times}(\hat{\Omega})\cos(2\psi)-F^{+}(\hat{\Omega})\sin(2\psi)] \nonumber\\&& \sin(2\pi f t+\phi_0-\Delta\Phi/2)\} \, ,
\label{TRtfull}
\end{eqnarray}
where we have defined a phase parameter as
\begin{equation}
\label{DeltaPhi}
\Delta \Phi = 2\pi f d_p(1-\cos\theta)/c\,\, .
\end{equation}
Note that $\Delta \Phi$ is the phase difference between pulsar terms and Earth terms for a non-evolving source. The underlying assumption in equation (\ref{TRtfull}) is that binary sources are strictly non-evolving and thus a single parameter $f$ is used to denote the GW frequency. In the following sections we will investigate how our analysis method performs for evolving sources (i.e., allowing the two terms to have different frequencies).

\section{The method}
\label{SVDmethod}
Building on coherent analysis algorithms for GW bursts used in ground-based interferometers \citep{WenSchutz05,Klimenko08,Sutton10}, \citet{Zhu15Single} presented a coherent method for the detection and localization of single-source GWs with PTAs. Here, we summarize its basic framework and extend it to include a coherent search for pulsar terms of non-evolving SMBHBs.

PTA data are traditionally presented in the time domain and in our method they need to be transformed to the frequency domain. For the evenly-sampled synthetic data sets used in this paper, this is done with a discrete Fourier Transform. For the irregularly sampled real data, their Fourier coefficients can be estimated with a maximum-likelihood approach \citep[e.g., section 3 in][]{PPTAcw14}. A similar treatment was applied to the GW background in \citet{Lentati13} where the signal is modeled as a number of frequency components.

Assuming Gaussian stationary noise and for a given source direction and GW frequency, timing residuals for an array of $N_{p}$ pulsars can be written in the simple matrix form
\begin{equation}
\label{data}
\mathbf{d}=\mathbf{F}\mathbf{A}+\mathbf{n},
\end{equation}
where $\mathbf{d}$, $\mathbf{n}$ and $\mathbf{F}$ are noise-weighted vectors or matrices that represent the data, noise and PTA's response respectively. The vector $\mathbf{A}$ represents the frequency-domain signal and is not noise-weighted. They are all complex vectors or matrices with detailed expressions given below.
\begin{equation}
\label{whitedata}
\mathbf{d}= \bb  d_{1}/\sqrt{S_{1}}\\d_{2}/\sqrt{S_{2}}\\ \vdots \\ d_{N_p}/\sqrt{S_{N_p}}\eb,
\mathbf{A} = \bb A_{+}\\ A_{\times}\eb,
\mathbf{n}=\bb  n_{1}/\sqrt{S_{1}}\\n_{2}/\sqrt{S_{2}}\\ \vdots \\ n_{N_p}/\sqrt{S_{N_p}}\eb,
\end{equation}
where $S_{j}$ is the one-sided noise power spectral density of the $j$-th pulsar, $A_{+}$ and $A_{\times}$ are full-term signals calculated using equations (\ref{Ap}-\ref{Ac}). The $j$-th row of matrix $\mathbf{F}$ is is given by
\begin{equation}
\label{whiteF}
\mathbf{F}_{j} = \frac{1-e^{-i\Delta \Phi_{j}}}{\sqrt{S_{j}}}\left[F^+_j\,\, F^\times_j \right] \, .
\end{equation}
Note that $\mathbf{F}$ depends on the pulsar position and GW source location but $\mathbf{A}$ does not, which greatly simplifies the analysis process. It is worth mentioning that our method works on a frequency-by-frequency basis. In practice $\mathbf{d}$ may have different length for different frequency since data span varies significantly for different pulsars. For example, the longest and shortest data span for the 49 pulsars included in the first IPTA data release is 27.1 and 4.5 years respectively, and 27 of them have a span over 10 years \citep{IPTA16}. However, this does not pose a problem for our method since Fourier components can be estimated at any given frequency with a maximum-likelihood or Bayesian approach. The application of our method to actual PTA data will be presented in a future work.

We apply the singular value decomposition to the response matrix $\mathbf{F}$
\begin{equation}
\label{svdF}
\mathbf{F}=\mathbf{U}\mathbf{S}\mathbf{V}^{\ast},
\mathbf{S} = \bb s_{1} &  0 \\ 0 & s_{2} \\ 0 & 0 \\ \vdots & \vdots \\ 0 & 0 \eb,
\end{equation}
where $\mathbf{U}$ and $\mathbf{V}$ are unitary matrices with dimensions of $N_p \times N_p$ and $2 \times 2$ respectively, the symbol $\ast$ denotes the conjugate transpose, and the rectangular ($N_p \times 2$) diagonal matrix $\mathbf{S}$ contains singular values of $\mathbf{F}$: $s_{1}$ and $s_{2}$, which are ranked such that $s_{1}\geqslant s_{2}\geqslant 0$. We then construct new data streams as follows:
\begin{equation}
\label{new-data}
\mathbf{\tilde{d}}=\mathbf{U}^{\ast}\mathbf{d}, \,\,  \mathbf{\tilde{A}}=\mathbf{V}^{\ast} \mathbf{A}, \,\,  \mathbf{\tilde{n}}=\mathbf{U}^{\ast}\mathbf{n} .
\end{equation}
One can see that the new data vector satisfies $\mathbf{\tilde{d}}= \mathbf{S}\mathbf{\tilde{A}}+\mathbf{\tilde{n}}$, and is specifically expressed as:
\begin{equation}
\label{new-data1}
\mathbf{\tilde{d}}=\bb  s_{1}\left(\mathbf{V}^{\ast} \mathbf{A}\right)_{1}+\mathbf{\tilde{n}}_{1} \\ s_{2}\left(\mathbf{V}^{\ast} \mathbf{A}\right)_{2}+\mathbf{\tilde{n}}_{2} \\ \mathbf{\tilde{n}}_{3} \\ \vdots \\ \mathbf{\tilde{n}}_{N_{p}}\eb .
\end{equation}
Here one can see that such an operation projects the data to two subspaces -- the first contain all information about GWs and the other one has null response to GWs. This is determined by the fact that in general relativity a GW has two independent polarization states. We refer the first two terms of $\mathbf{\tilde{d}}$ to \emph{signal streams} and the remaining to \emph{null streams}.

The maximum likelihood estimator for $\mathbf{A}$ is
\begin{equation}
\label{Aest}
\hat{\mathbf{A}}=\bar{\mathbf{F}}\mathbf{d},
\end{equation}
where $\bar{\mathbf{F}}$ is the pseudoinverse of $\mathbf{F}$, which can be obtained through its singular value decomposition
\begin{equation}
\label{Fbar}
\bar{\mathbf{F}}=\mathbf{V}\bar{\mathbf{S}}\mathbf{U}^{\ast}, \,\bar{\mathbf{S}} = \bb 1/s_{1} &  0 & 0 & \hdots & 0 \\ 0 & 1/s_{2} & 0 & \hdots & 0 \eb.
\end{equation}

We use the following test statistic for monochromatic GWs:
\begin{equation}
\label{DS-mon}
\mathcal{P}= \sum^{2}_{j=1} |\mathbf{\tilde{d}}_j|^{2}.
\end{equation}
It is worth mentioning that this method is optimal under the Neyman-Pearson criterion for monochromatic signals with known frequencies \citep{FlaHughes98,ExcessPower01,Sutton10}.

In the absence of GW signals, $\mathcal{P}$ follows a $\chi^2$ distribution with four degrees of freedom; when signals are present it follows a noncentral $\chi^2$ distribution with a noncentrality parameter $\rho^2$
\begin{equation}
\label{DS_rho}
\langle\mathcal{P}\rangle= 4 + \rho^{2},
\end{equation}
where the brackets $\langle ... \rangle$ denote the ensemble average of the random noise process. Note that $\rho$ is the expected signal-to-noise ratio and $\rho^2$ equals the statistic calculated for noiseless data.

\subsection{Maximization over intrinsic parameters}
The calculation of our test statistic requires information of source sky location ($\alpha$ and $\delta$), GW frequency ($f$) and phase parameters $\Delta \Phi_{j}$. For a blind all-sky search, none of these is known \emph{a priori}. Therefore, a search is performed over the allowed parameter space for the maximum test statistic, which is called the GLRT (generalized likelihood ratio test) statistic. This will incur a false-alarm penalty that needs to be accounted for carefully. The false alarm probability of a measured statistic $\mathcal{P}_{0}$ is the probability that $\mathcal{P}$ exceeds the measured value for noise-only data. It is given by $1-[\mathrm{CDF}(\mathcal{P}_{\rm{max}};\chi^{2}_{4})]^{N_{\rm{trial}}}$ for the GLRT statistic ($\mathcal{P}_{\rm{max}}$) found in the search, where $N_{\rm{trial}}$ is the trials factor defined as the number of independent cells in the searched parameter space. While it may be difficult to calculate $N_{\rm{trial}}$ analytically, one can use simulations to obtain the empirical cumulative distribution of $\mathcal{P}_{\rm{max}}$ from searches performed in noise-only data sets.

The parameter space for a full-signal search is $3+N_p$ dimensional, with $N_p$ ranging from about 20 to 40 for current PTAs. While a grid-based brute-force search is sufficient and easy to implement for Earth terms (over a 3-dimensional space), more intelligent search algorithms are required for the search of the full signal. In this work we make use of the Particle Swarm Optimization (PSO) algorithm to deal with the numerical maximization problem. PSO is a population-based stochastic optimization method developed by \citet{eberhart95}. It has been applied to a wide range of computational problems. In the context of GW data analysis, \citet{YanWang10PRD} first used it for the detection and estimation of binary inspirals with ground-based detectors. More recently, \citet{YanWang14,YanWang15} applied it to the problem of detection and parameter estimation of continuous GWs using PTAs. These two papers dealt with exactly the same problem as this work with the main difference lying in the construction of test statistics.

Very briefly, PSO works as follows. Initially one gives a population of \emph{particles} whose positions in the parameter space are determined randomly. Each particle represents a candidate solution and one expects to find the best solution with respect to a given measure of quality (which in our case is the test statistic $\mathcal{P}$). These particles move around the parameter space with the movement being guided by both their personal best positions and the global best position (as found by all particles in the past). In this way, candidate solutions are iteratively improved and finally the global best at termination of the iterations provides the best solution found by PSO. It is interesting to note that the evolution of the particle swarm mimics in some ways the social behavior of bird flocking or fish schooling. We refer interested readers to the original paper by \citet{eberhart95} and to WMJ15 for detailed description of the algorithm.

For this work, we use the same set-up as described in section 3.3 of WMJ15. We adjust two key PSO parameters (a) the number of particles and (b) the total number of iterations to increase the probability of successful convergence. For simulated data used in the following section, a convergence is declared to be successful if (1) the maximum statistic found by PSO is no smaller than the value returned by using true signal parameters and (2) multiple independent runs of PSO return essentially the same test statistic (e.g., differ by no more than $0.1\%$).

\section{Results}
\label{sec-results}
In this section, we apply our analysis techniques to simulated PTA data sets. We consider 30 IPTA pulsars as listed in Table \ref{tb:rmsIPTApsr}. The rms timing residuals (white-noise levels) are arbitrarily set but broadly represent the actual measurements of current PTA data sets \citep{PPTA2013,NANO15data,epta16data,IPTA16}. Our fiducial array consists of 12 ``best" pulsars that have rms white noise levels below 300 ns, while the 30-pulsar array is used for discussions concerning the computational cost.

For all simulations we assume a 15-year data span and 2-week cadence. Data sets are produced using independent white noise realizations and (when applicable) signals expected from SMBHBs in circular orbits. The signal frequency is set at 10 nHz because (i) it has been shown to be around the most sensitive frequency band of current PTAs \citep[e.g.,][]{PPTAcw14} and (ii) it is least affected by the fitting to a pulsar timing model. The fitting process is known to significantly reduce a PTA's sensitivity to GWs (i) at the lowest frequencies $\sim 1/T_{\rm{obs}}$ (because of the fit for pulsar spin period and its first time derivative) and (ii) in two narrow bands centred around 1 $\rm{yr}^{-1}$ (the fit for pulsar positions and proper motions) and 2 $\rm{yr}^{-1}$ (the fit for pulsar parallax). We therefore ignore the timing-model fit in our analysis.

\begin{table}
\begin{center}
\caption{The rms timing residual levels and distances for the 30 IPTA pulsars considered in this work. Our fiducial array consists of 12 ``best" pulsars that have rms noise below 300 ns. Values of pulsar distances are taken from the \href{http://www.atnf.csiro.au/research/pulsar/psrcat/}{ATNF Pulsar Catalogue} \citep{ATNF05Pulsar}. Note that distances are poorly known for most of pulsars listed here.}
\label{tb:rmsIPTApsr}
\vspace{2mm}
\begin{tabular}{lcclcc}
\hline
\hline
            &  Res.   & $d_p$ &   &  Res.  & $d_p$ \\
  PSR       &  (ns)  & (kpc) & PSR   & (ns)  & (kpc) \\
\hline
  J0437$-$4715 &  58  & 0.16  & J1600$-$3053  & 202 & 2.40  \\
  J1640$+$2224 &  158 & 1.19  & J1713$+$0747  & 116 & 1.05  \\
  J1741$+$1351 &  233 & 0.93  & J1744$-$1134  & 203 & 0.42  \\
  J1909$-$3744 &  102 & 1.26  & J1939$+$2134  & 104 & 5.00  \\
  J2017$+$0603 &  238 & 1.32  & J2043$+$1711  & 170 & 1.13  \\
  J2241$-$5236 &  300 & 0.68  & J2317$+$1439  & 267 & 1.89  \\
  J0023$+$0923 &  320 & 0.95  & J0030$+$0451  & 723 & 0.28 \\
  J0613$-$0200 &  592 & 0.90  & J1017$-$7156  & 500 & 0.26 \\
  J1024$-$0719 &  846 & 0.49  & J1446$-$4701  & 500 & 2.03 \\
  J1614$-$2230 &  336 & 1.77  & J1738$+$0333  & 316 & 1.47 \\
  J1832$-$0836 &  577 & 1.40  & J1853$+$1303  & 369 & 1.60 \\
  J1857$+$0943 &  505 & 0.90  & J1911$+$1347  & 500 & 1.60 \\
  J1918$-$0642 &  547 & 1.40  & J1923$+$2515  & 535 & 0.99 \\
  J2010$-$1323 &  733 & 1.29  & J2129$-$5721  & 880 & 0.40 \\
  J2145$-$0750 &  535 & 0.57  & J2214$+$3000  & 399 & 1.32 \\
\hline
\hline
\end{tabular}
\end{center}
\end{table}

\subsection{Examples for different signal-to-noise ratios}
\label{Examps}
To demonstrate the effectiveness of our method, we consider a circular binary system located in the Virgo cluster ($d_L=$ 16.5 Mpc) characterized by the following parameters: $f=10$ nHz, $\cos\iota=0.88$, $\psi=0.5$ rad, $\phi_{0}=2.89$ rad, $(\alpha,\,\delta)=$(3.2594 rad, 0.2219 rad). It has been suggested that the Virgo cluster may represent a GW hotspot for PTA searches \citep{SimonGWhot14}. The value of $h_0$ is scaled to suit the desired signal-to-noise ratio defined as:
\begin{equation}
\label{netSNR}
\rho^{2} = \sum_{j=1}^{N_{p}}\rho_{j}^{2} = \sum_{j=1}^{N_{p}}\sum_{i=1}^{N}\left[\frac{s(t_i)}{\sigma_{j}}\right]^{2}\, ,
\end{equation}
where $\rho_{j}$ is the individual signal-to-noise ratio for each pulsar, $N$ is the total number of data points ($N=392$ for our simulations), $\sigma$ is the noise rms listed in Table \ref{tb:rmsIPTApsr}.

Here we consider the following three sets of signals
\begin{enumerate}
\item $M_{c}=8.77 \times 10^{8} M_{\odot}$ ($h_0=1.34 \times 10^{-14}$), $\rho=100$;
\item $M_{c}=4.26 \times 10^{8} M_{\odot}$ ($h_0=4.03 \times 10^{-15}$), $\rho=30$;
\item $M_{c}=1.93 \times 10^{8} M_{\odot}$ ($h_0=1.07 \times 10^{-15}$), $\rho=8$.
\end{enumerate}
These correspond to the strong, moderate and weak signal cases respectively. It is worth mentioning that the presence of a strong signal as in case (i) in the Virgo cluster has already been ruled out by current PTA single-source limits \citep{PPTAcw14,EPTAcw15}. We use it here only for the purpose of testing the analysis method. To facilitate later discussions, we show in Table \ref{tb:rhoTheta} values of $\rho_{j}$ for case (i) as well as the angular separation between the GW source and each pulsar.

For case (i) which has the highest chirp mass, the maximum frequency drift between pulsar terms and Earth terms is 1.08 nHz for J1939$+$2134 (which is not surprising because of its significantly larger pulsar distance). This implies that it is appropriate to assume a non-evolving signal model. For each case, we add the simulated signal to 500 independent realizations of Gaussian noise, allowing the conventional Frequentist error estimates for signal parameters to be obtained. We also compare parameter estimation results with that from an Earth-term search \citep{Zhu15Single}.

\begin{table}
\begin{center}
\caption{Values of individual signal-to-noise ratio ($\rho_{j}$) for the strong signal injection ($\rho=100$). Also listed is the angular separation ($\theta$) between the GW source and each pulsar.}\label{tb:rhoTheta}
\vspace{2mm}
\begin{tabular}{lcclcc}
\hline
\hline
  PSR       &  $\rho_{j}$  & $\theta$ ($^{\circ}$) & PSR   & $\rho_{j}$  & $\theta$ ($^{\circ}$) \\
\hline
  J0437$-$4715 &  47.9  & 118  & J1600$-$3053  & 35.8 & 67  \\
  J1640$+$2224 &  17.2 & 61  & J1713$+$0747  & 56.5 & 71  \\
  J1741$+$1351 &  27.4 & 76  & J1744$-$1134  & 17.6 & 82  \\
  J1909$-$3744 &  29.1 & 106  & J1939$+$2134  & 31.1 & 102  \\
  J2017$+$0603 &  12.0 & 115  & J2043$+$1711  & 6.2 & 117  \\
  J2241$-$5236 &  1.8 & 135  & J2317$+$1439  & 2.1 & 148  \\
\hline
\hline
\end{tabular}
\end{center}
\end{table}

\subsubsection{Strong signal}

\begin{figure}
\includegraphics[width=0.45\textwidth]{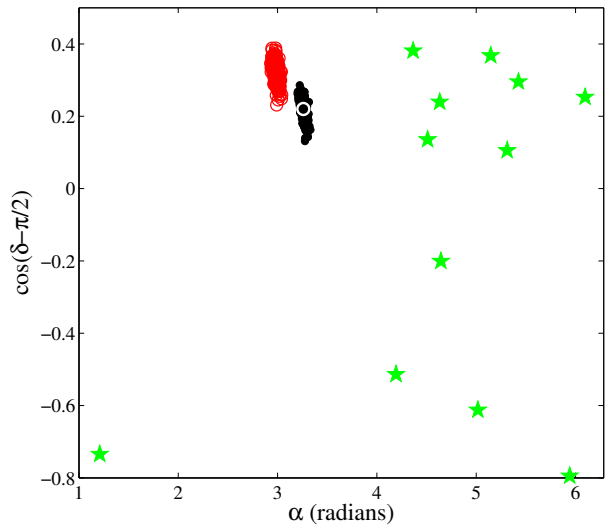}
\caption{A scatter plot of the estimated source sky location for a strong signal ($\rho=100$) injected to 500 noise realizations. We compare the results from a full-signal search (black dots) with that of an Earth-term search (red circles). The true source location is marked by a white ``$\circ$". Sky locations of 12 IPTA pulsars are marked with green ``$\star$".}
\label{fig:SkyLoc100}
\end{figure}

Figure \ref{fig:SkyLoc100} shows the sky localization results for the strong signal case ($\rho=100$). One can see that the source is successfully localized on the sky (within $\lesssim5$ degrees of the true location) with the current method, while an Earth-term search returns systematically biased results. Such a bias is caused by the presence of pulsar terms which are not accounted for in the search. The signal is so strong that the noise-induced scatter does not even overlap with the true sky location for the Earth-term search. On the other hand, we note that the overall spread (equivalent to the 3-sigma contour) of the estimated locations is comparable for both searches, enclosing an area of $\sim 40$ deg$^{2}$.

For such a strong signal, the GW frequency is estimated extremely well for both searches -- for all noise realizations the frequency is localized within the two closest bins around 10 nHz. Since we oversample the frequency by a factor of 32 in the Fourier transform so that the frequency resolution is effectively $1/(32\hspace{0.5mm}T_{\rm{obs}})$. This lead to an accuracy of frequency estimation $\sim 0.01/T_{\rm{obs}}$ in this case.

\begin{figure*}
\centerline{\includegraphics[width=0.95\textwidth]{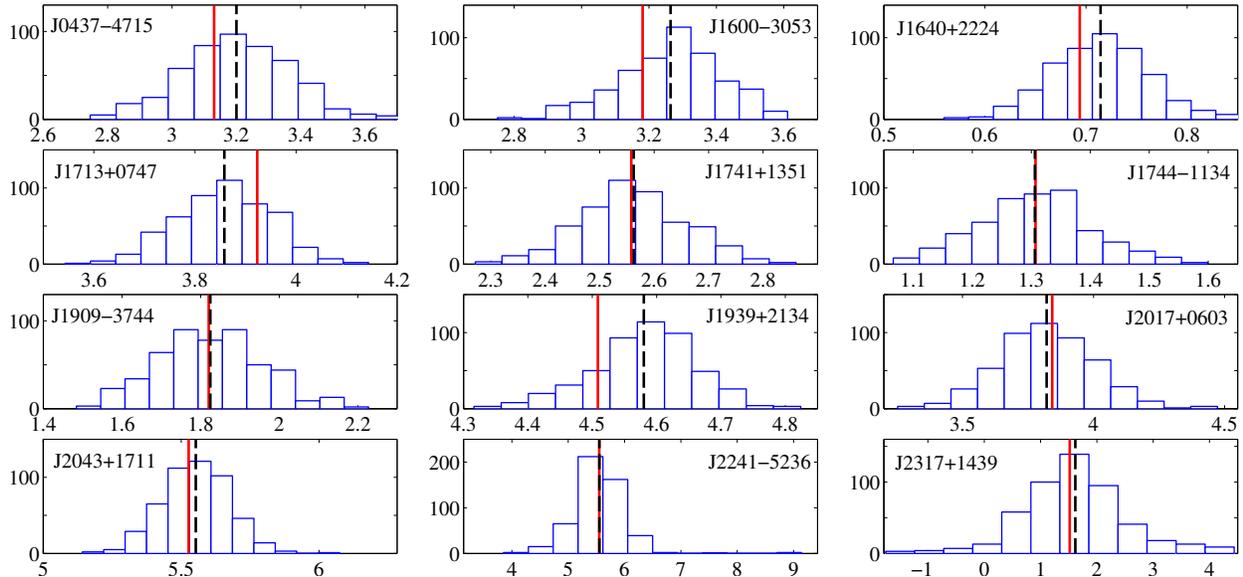}}
\caption{Histograms of phase parameters ($\Delta \Phi_{j}$, in radians) from an analysis of a strong signal ($\rho=100$) injected to 500 noise realizations. The black dash and red solid line marks the estimated mean and true parameter respectively.}
\label{fig:Pha12p100}
\end{figure*}

Figure \ref{fig:Pha12p100} shows the distribution of estimated phase parameters ($\Delta \Phi$) for all 500 noise realizations, with the true and mean values also indicated. It is clear that for all pulsars $\Delta \Phi$ is successfully recovered and the injected value lies within the one-sigma credible interval. The standard deviations ($\sigma_{\Delta \Phi}$) range from 0.047 rad (for J1640$+$2224, which has the smallest angular separation from the GW source) to 0.92 rad (for J2317$+$1439 which has the largest angular separation from the GW source). For most pulsars $\Delta \Phi$ can be measured with a precision better than $\sim 0.15$ rad except J2241$-$5236 and J2317$+$1439, both of which have low $\rho_{j}$ and large $\theta$ (see Table \ref{tb:rhoTheta}). Note that in GW analysis it is the modulo of $\Delta \Phi$ with respect to $2\pi$ that can be measured. The search range for $\Delta \Phi_{j}$ is from 0 to $2\pi$. The distribution shown in Figure \ref{fig:Pha12p100} (and similar figures hereafter) has been wrapped to an appropriate range by projecting some estimated values to $\Delta \Phi+2\pi$ or $\Delta \Phi-2\pi$.

\subsubsection{Moderate signal}
\begin{figure*}
\centerline{\includegraphics[width=0.95\textwidth]{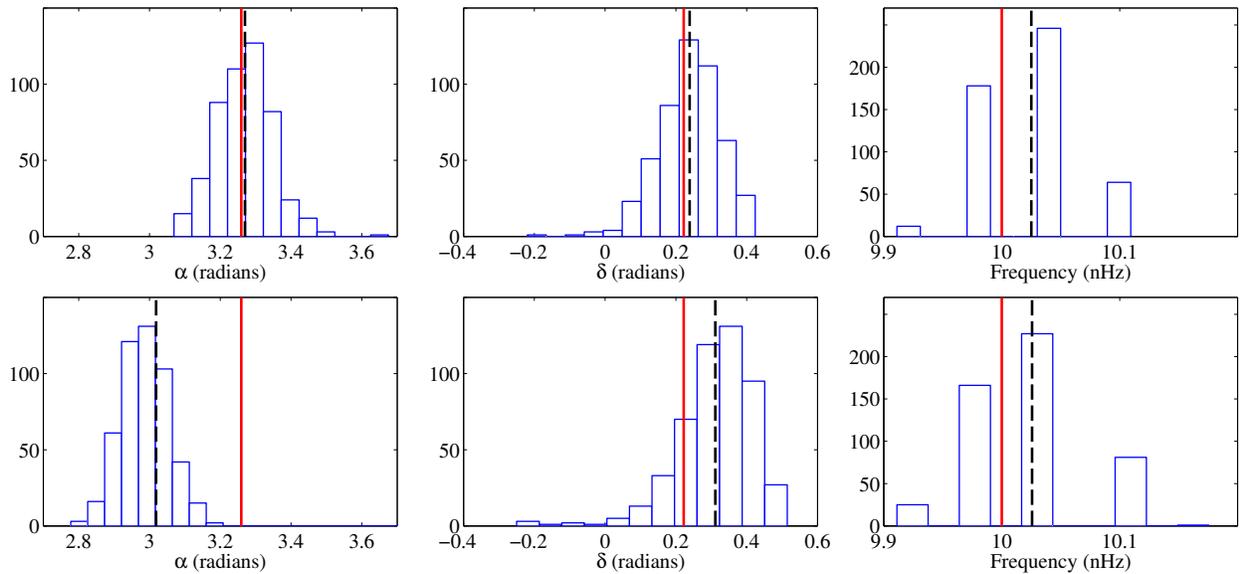}}
\caption{Distribution of estimated source sky location ($\alpha$ and $\delta$) and GW frequency for the moderate signal case ($\rho=30$). We performed both a full-signal search (upper row) and an Earth-term search (lower row) in 500 noise realizations. For each plot the black dash and red solid line marks the mean estimation and injected value of each parameter respectively.}
\label{fig:RADecF30}
\end{figure*}

Figure \ref{fig:RADecF30} shows histograms of estimated sky location ($\alpha$ and $\delta$) and GW frequency for the moderate signal case ($\rho=30$). We again compare the results from the current method (upper panels in the figure) with those from an Earth-term search (lower panels). For a full-signal search, all three parameters are comfortably recovered, with the injected values located within their respective one-sigma intervals. The standard deviations are 6.6$^{\circ}$ and 5.7$^{\circ}$ for $\alpha$ and $\delta$ respectively. Similar to the strong signal case, estimation of $\alpha$ from the Earth-term search is completely off its true value. The situation gets improved for $\delta$, since the injected value is within the one-sigma confidence interval as found in the Earth-term search. On the other hand, the frequency estimation is comparable in both searches with a standard deviation of about 0.05 nHz.

\begin{figure*}
\centerline{\includegraphics[width=0.95\textwidth]{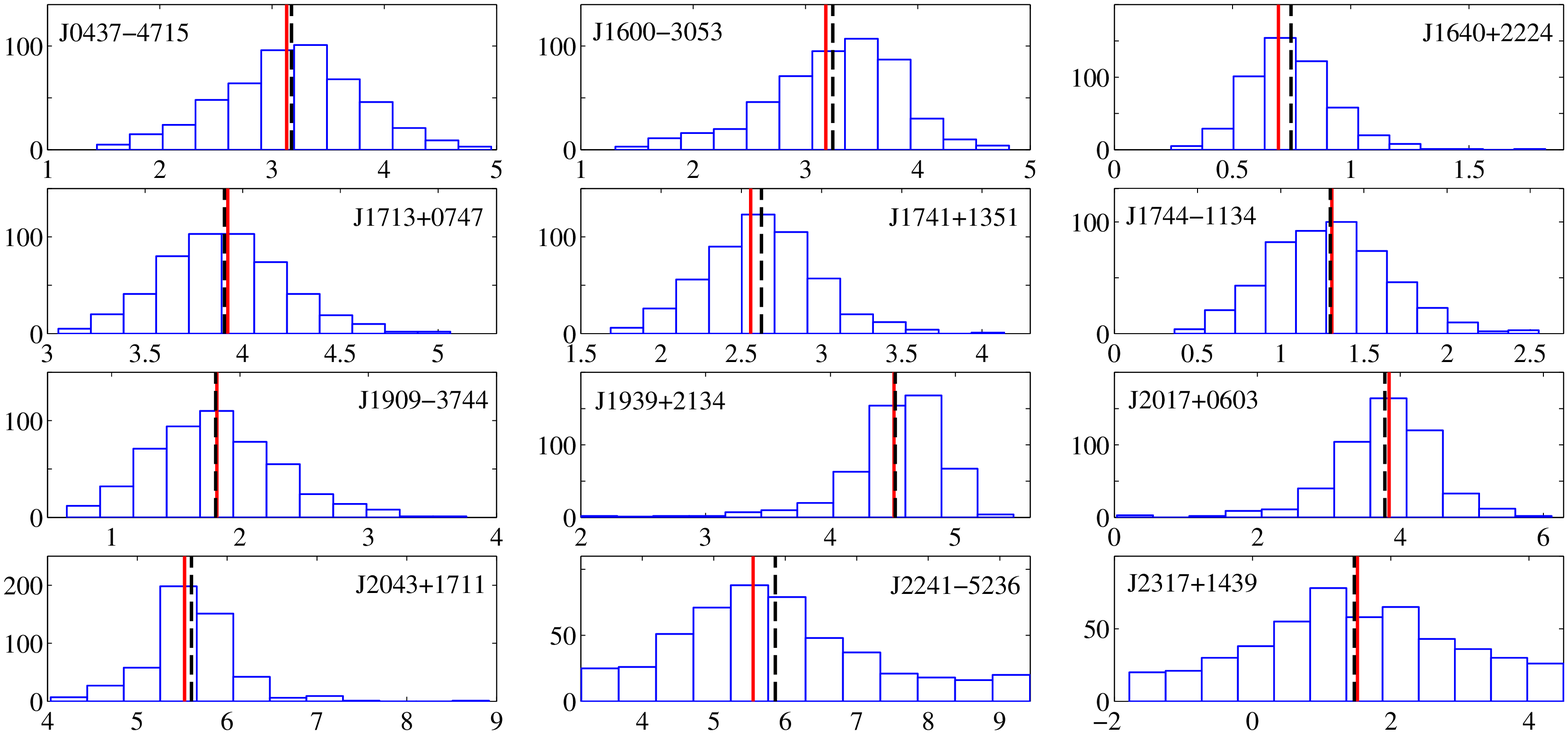}}
\caption{As Figure \ref{fig:Pha12p100} but for a moderate signal case ($\rho=30$).}
\label{fig:Pha12p30}
\end{figure*}

Figure \ref{fig:Pha12p30} shows the histograms for the phase parameters $\Delta \Phi$. It is clear that a Gaussian distribution can be seen and the true value is recovered within the one-sigma confidence interval for all pulsars. The standard deviation is $\lesssim 0.5$ rad for most pulsars. For J2241$-$5236 and J2317$+$1439, reasonably good estimation of $\Delta \Phi$ is still possible despite the small individual signal-to-noise ratio $\sim 0.5$. When comparing with Figure \ref{fig:Pha12p100}, the scaling of $\sigma_{\Delta \Phi}$ is consistent with $1/\rho$.

\begin{figure}
\includegraphics[width=0.45\textwidth]{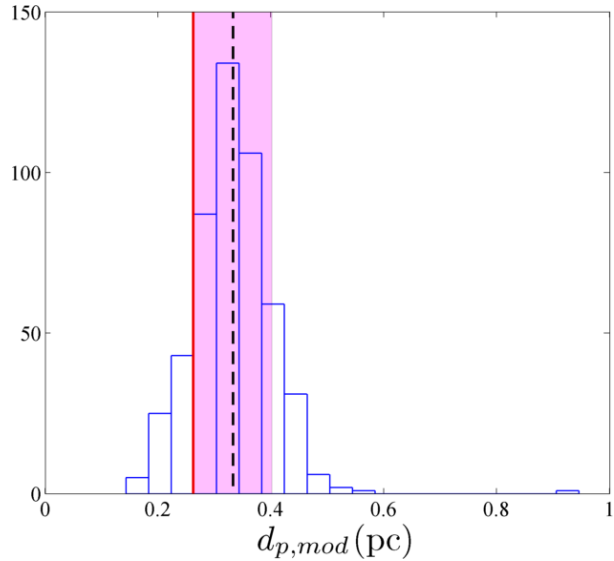}
\caption{Distribution of the ``reduced" distance ($d_{p,mod}$) for J0437$-$4715 from our analysis of a moderate GW signal ($\rho=30$) injected to 500 noise realizations. The black dash and red solid line marks the mean and true value respectively. The shaded region indicates the one-sigma confidence interval.}
\label{fig:dpJ0437}
\end{figure}

Precise estimates of $\alpha$, $\delta$, $f$ and $\Delta \Phi$ may provide some useful constraints on pulsar distance. We explore this possibility in the moderate signal case. For each noise realization, we combine those estimated parameters to infer the following ``reduced" distance
\begin{equation}
d_{p,mod} = \frac{c \Delta \Phi_{mod}}{2\pi f (1-\cos\theta)}\, ,
\label{dpmod}
\end{equation}
where $\Delta \Phi_{mod}$ is the modulo of $\Delta \Phi$ with respect to $2\pi$. In order for this quantity to be useful, the pulsar distance must already be known precisely, with its entire uncertainty range being smaller than $\lambda_{GW}/(1-\cos\theta)$, e.g., $<0.66$ pc for J0437$-$4715 (i.e., the pulsar with the most precisely measured distance so far) for the GW source considered here. Current best measurement for J0437$-$4715 \citep{Reardon16} has a one-sigma range of $0.5$ pc, which is a factor of $\sim 3$ away from the required precision. But it is already useful for lower $f$ and/or smaller $\theta$. For example, the Reardon et al. measurement could be useful for pulsar-term searches when $\theta<70^{\circ}\,(107^{\circ})$ at $f=10\,(5)$ nHz.

Nevertheless, we show in Figure \ref{fig:dpJ0437} the distribution of $d_{p,mod}$ obtained from our analysis for J0437$-$4715. With uncertainties in the estimation of all relevant parameters taken into account, the true value of $d_{p,mod}$ is recovered within the one-sigma ($\pm 0.07$ pc) confidence interval.

\subsubsection{Weak signal}

\begin{figure*}
\centerline{\includegraphics[width=0.95\textwidth]{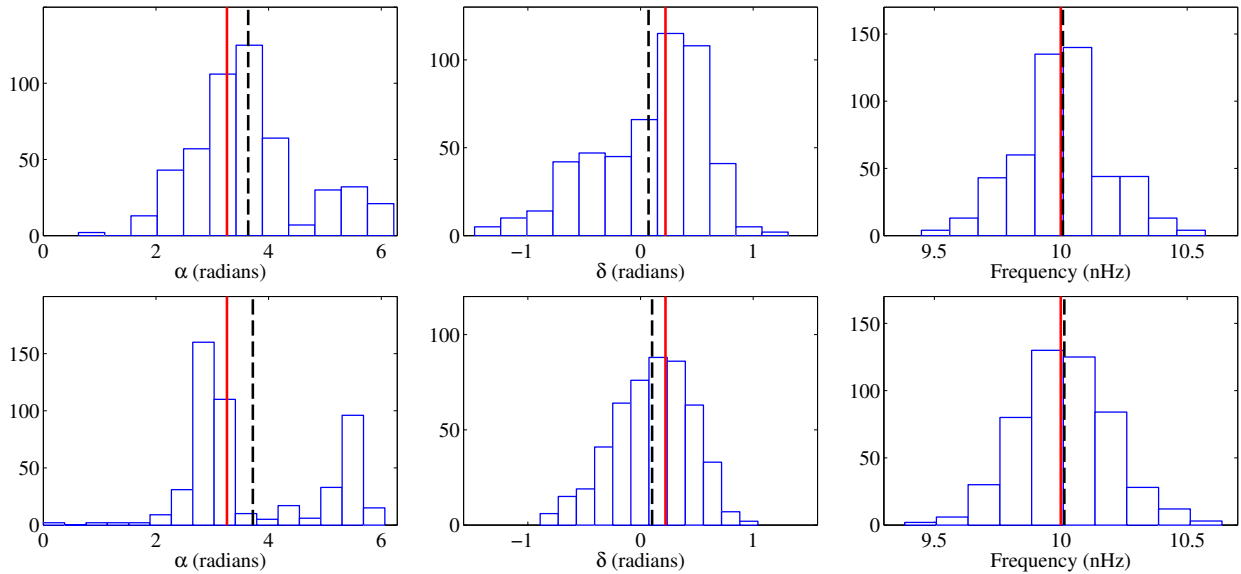}}
\caption{As Figure \ref{fig:RADecF30} but for a weak signal case ($\rho=8$).}
\label{fig:RADecF8}
\end{figure*}

Figure \ref{fig:RADecF8} shows histograms for the estimated sky location ($\alpha$ and $\delta$) and GW frequency for the weak signal case ($\rho=8$). As in the previous two cases, the GW frequency can be recovered very well for both searches with a standard deviation of about 0.2 nHz. However, unlike the previous two cases where the sky localization of the Earth-term search is completely off the true location, the injected values of both $\alpha$ and $\delta$ are within the one-sigma confidence intervals as found in the Earth-term search. By visual inspection, one can see that the Earth-term search gives similar estimates of $\delta$ but notably worse results for $\alpha$. After performing a two-sample Kolmogorov-Smirnov test, we find that distributions of estimated $\alpha$ and $\delta$ are significantly different for the two methods.

\begin{figure*}
\centerline{\includegraphics[width=0.95\textwidth]{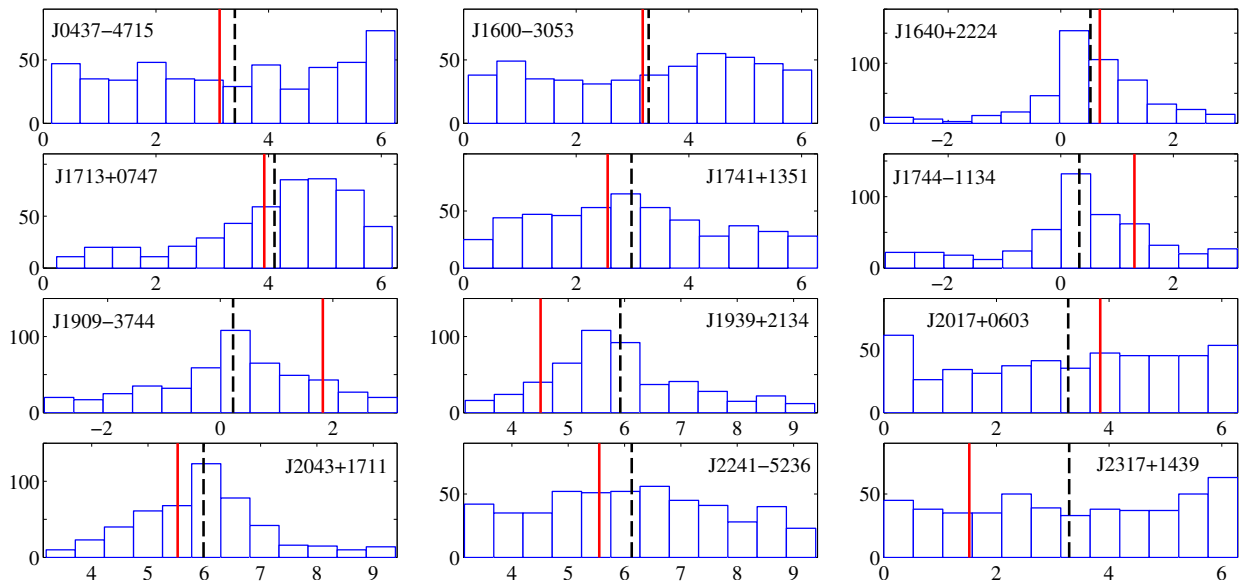}}
\caption{As Figure \ref{fig:Pha12p100} but for a weak signal case ($\rho=8$).}
\label{fig:Pha12p8}
\end{figure*}

Figure \ref{fig:Pha12p8} shows the distribution of $\Delta \Phi$ for each pulsar. In contrast to previous two cases, for most pulsars either the resulting distribution is broadly uniform or the true parameter lies outside the one-sigma confidence interval. The distribution occupies the entire $2\pi$ range for all pulsars. Still, reasonably good estimation is obtained for a few pulsars despite the small individual signal-to-noise ratios. In particular, for J1640$+$2224 which is closest to the GW source on the sky, $\Delta \Phi$ is successfully recovered with a standard deviation of $\sim1$ rad.

\subsection{On the detection probability}
Here we quantify how the coherent inclusion of pulsar terms in the single-source GW searches will boost the chance of a detection compared with an Earth-term search. We do this by applying our analysis techniques to a large set of simulations. For a fixed signal-to-noise ratio $\rho=8$, we perform both a full-signal search and an Earth-term search for the following cases:
\begin{enumerate}
\item non-evolving, signals generated with equation (\ref{TRtfull});
\item weakly evolving, $M_{c}=10^{9} M_{\odot}$;
\item evolving, $M_{c}=3 \times 10^{9} M_{\odot}$.
\end{enumerate}
For the latter two cases, the frequency of pulsar terms is determined with equations (\ref{TpTe}) and (\ref{GW-freq}) and the signals are generated as the sum of two sinusoids with different frequencies. In all three cases, the Earth-term frequency is 10 nHz, the source parameters $\{\alpha,\delta,\cos\iota,\psi,\phi_{0}\}$ are randomized, and $h_0$ (or $d_L$ in the case of given chirp mass) is scaled such that $\rho=8$. For a uniform distribution of source sky locations, the frequency drift between pulsar terms and Earth terms is $< 1/T_{\rm{obs}}$ for the weakly-evolving case and $\gtrsim 1/T_{\rm{obs}}$ for the evolving case.

We perform searches on $2\times 10^{3}$ synthetic data sets that consist of independent noise realizations and simulated signals. For each data set we record the GLRT statistic found by a full-signal search ($\mathcal{P}$) described in Section \ref{SVDmethod} or an Earth-term search ($\mathcal{P}_{\rm{et}}$) presented in \citet{Zhu15Single}. We set the detection threshold for a given false alarm probability using the empirical distribution of the GLRT statistic returned from searches performed in a large number (e.g., $10^{4}$ is used in this work) of noise-only data sets. The settings (e.g., the number of particles and iterations) for our PSO search algorithm are kept the same for simulations with and without signal injections.

\begin{figure}
\includegraphics[width=0.45\textwidth]{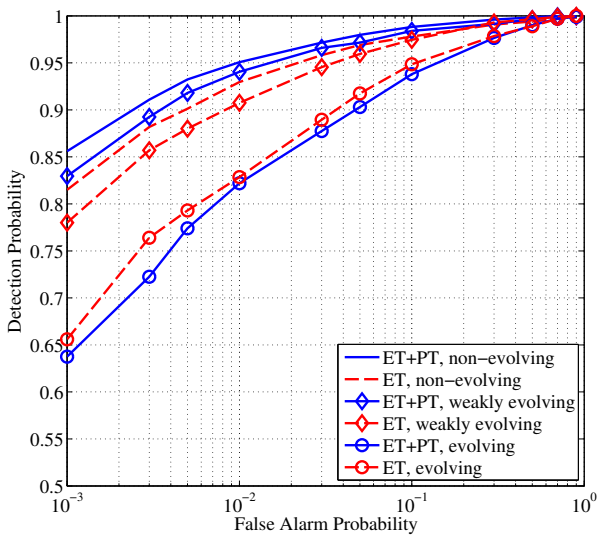}
\caption{Receiver-operator characteristic curves (detection probabilities as a function of false alarm probability) for $\rho=8$. Here we compare the detection method that targets both Earth-term (ET) and pulsar-term (PT) signals (solid blue lines) and a method that targets only ET signals (red dash lines). Three cases of frequency evolution (between PT and ET) are considered: (1) non-evolving (lines without markers), (2) weakly evolving (marked with $\diamond$) and (3) evolving (marked with $\circ$). Because of the large number ($>2000$) of simulations performed, the one-sigma fluctuation of these curves is estimated (from a binomial statistic) to be $<1\%$.}
\label{fig:FAR_DP}
\end{figure}

Figure \ref{fig:FAR_DP} shows the performance of the two test statistics ($\mathcal{P}$ and $\mathcal{P}_{\rm{et}}$) for different signal models. At a given false alarm probability of $10^{-3}$, $\mathcal{P}$ offers slightly higher detection probabilities (by $5\%$) than $\mathcal{P}_{\rm{et}}$ for both the non-evolving and weakly-evolving scenarios. Such a modest (rather than dramatic) enhancement can be interpreted as follows. A full-signal search is beneficial because it picks up some extra signal power that is thrown away in the Earth-term search. Note that for signals simulated in this section with $\rho=8$, the Earth-term signal-to-noise ratio is $\sim 6$. But the benefit is compromised by the false-alarm penalty incurred by the higher-dimensional search for unknown intrinsic source parameters. For our fiducial 12-pulsar array, the search dimension increases from 3 for $\mathcal{P}_{\rm{et}}$ to 15 for $\mathcal{P}$.

Now let us look at the curves of $\mathcal{P}$ and $\mathcal{P}_{\rm{et}}$ separately in Figure \ref{fig:FAR_DP}. The performance of $\mathcal{P}$ drops by only $3\%$ in the weakly-evolving case compared with the non-evolving case. This implies that $\mathcal{P}$ is a nearly optimal detection method for weakly evolving SMBHBs. Additionally, identical results of parameter estimation are obtained if we reanalyse examples shown in section \ref{Examps} for weakly-evolving signals. On the other hand, the detection probability of $\mathcal{P}$ decreases significantly for the evolving case, but interesting to note that it is still comparable to that of $\mathcal{P}_{\rm{et}}$. In this regard, alternative analysis methods are required for the detection and parameter estimation of evolving SMBHBs \citep{Taylor14}. The drop in detection probability of $\mathcal{P}_{\rm{et}}$ from non-evolving to evolving cases is due to the decrease of Earth-term signal-to-noise ratio. Pulsar terms are at lower frequencies and thus have higher amplitudes, implying that they contribute more to the total signal-to-noise ratio (which is kept the same for all cases).

\subsection{On the sky localization and computational cost}
\begin{figure*}
\includegraphics[width=0.95\textwidth]{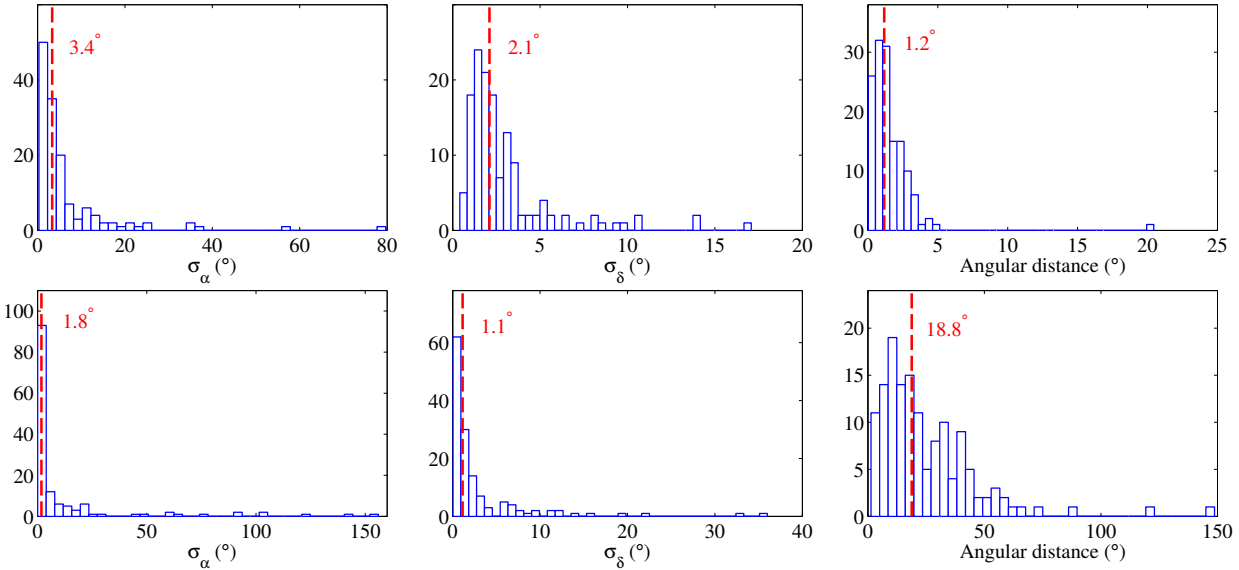}
\caption{For 140 uniformly distributed GW source sky locations, we analyze 200 noise realizations with a strong non-evolving signal ($\rho=100$) injected. We show the distribution of the standard deviation of $\alpha$ and $\delta$, and the angular distance between the true source sky location and the mean location found in the search. The upper panels are for a full-signal search and the low panels are for an Earth-term search. The red dash line marks the median value for the 140 sky locations.}
\label{fig:LocEPTall}
\end{figure*}

To have a more complete picture on how important a full-signal search is for obtaining reliable sky localization of non-evolving SMBHBs, we repeat the analysis in section \ref{Examps} for 140 uniformly distributed source sky locations. For each source location, we inject a strong signal ($\rho=100$) to 200 noise realizations. We analyze the data with both statistics $\mathcal{P}$ and $\mathcal{P}_{\rm{et}}$. We found that the true source location lies within the one-sigma confidence interval for all 140 sky locations considered for $\mathcal{P}$, while this happens only for $22\%$ of the sky using $\mathcal{P}_{\rm{et}}$.

Figure \ref{fig:LocEPTall} shows the distribution of the standard deviation of estimated $\alpha$ and $\delta$, and the angular distance between the true source sky location and the mean location found in the analysis. It is interesting to note that $\sigma_{\alpha}$ and $\sigma_{\delta}$ of $\mathcal{P}$ are larger than those of $\mathcal{P}_{\rm{et}}$ for a typical sky location, but their ranges are significantly smaller. The estimation of $\alpha$ is worse than that of $\delta$ for both methods. This is due to the inhomogeneous distribution of the pulsar sky locations (see Figure \ref{fig:SkyLoc100}). The long tails shown in the distributions of Figure \ref{fig:LocEPTall} can also be attributed to this fact since all of them occur in the region of $2\lesssim \alpha \lesssim \pi$. Overall, one can see that accounting for pulsar terms is critical for unbiased sky localization.

Regarding computational cost, we found that a single set-up of PSO does not work for all source sky locations. Significantly larger number of particles and iterations are required for some than the others. For a typical sky location and a simulated data set with $\rho=10$, detection and parameter estimation with our method can be finished within 1.5 and 4 minutes on a single processor core for a 12- and 30-pulsar array respectively. Given the same data span, observing cadence and signal strength, it takes 6.7 and 89 minutes for the method presented in \citet{YanWang14} and WMJ15 respectively for a 9-pulsar array data set. The computational cost for similar analysis is orders of magnitude higher for Bayesian methods. For example, \citet{Taylor14} reported that about 45 minutes on 48 computational cores are required for the analysis of a similar data set.

\section{Conclusions}
\label{Conclu}
Current PTA experiments have achieved unprecedented sensitivity to GWs in the nHz frequency band. They may open up a new window onto the nanohertz gravitational Universe in the near future. While the first detection is likely to be a stochastic background from numerous inspiralling SMBHBs distributed throughout the Universe, detected single events will provide much richer information about their astrophysical sources. Furthermore, detections of individual SMBHBs in the PTA band may serve as triggers for follow-up observations through the electromagnetic emission \citep{Tanaka12,Sesana12Multi,SarahCQG13} and complement observations using future space-based GW detectors such as eLISA \citep{eLISA12CQG}. A key to the practice of multimessenger astronomy is reliable sky localization of GW sources.

In this paper we have described a coherent method for the detection and sky localization of continuous GWs from non-evolving binaries. This method extends the Earth-term-based approaches presented in \citet{Zhu15Single} to coherently include the contribution of pulsar terms. We have shown that it is fast and reliable for the detection, sky localization and frequency estimation. Unlike previous studies, we find that the pulsar phase parameters can be estimated with relatively high accuracy. This allows the possibility to improve pulsar distance measurements through GW detections. Moreover, we find that our method is about a factor of ten faster than previously published Frequentist methods and orders of magnitude faster than existing Bayesian methods. The new method can be applied to a full-scale timing array that consists of $\gtrsim 30$ pulsars without causing any computational overload.

By comparing with an Earth-term searching method, we show that a coherent inclusion of pulsar terms can only improve the detection probability by $5\%$. We also find that 1) source localization from an Earth-term search is biased and accounting for pulsar terms is critical for removing such a bias; 2) the one-sigma error region of source localization is comparable for both searches, typically enclosing an area of more than hundreds of deg$^2$ for a signal-to-noise ratio of 10. Because pulsar distances are poorly constrained with uncertainties much larger than the wavelength of nanohertz GWs, we choose to fit for the pulsar phase parameters. Significantly better angular resolution may become achievable if distances to some pulsars are constrained within a reduced gravitational wavelength, which will be investigated in a future study.

In summary,
\begin{itemize}
\item our method allows the pulsar term to be accounted for when searching for continuous GWs. We only consider non-evolving SMBHBs for which there exists an arbitrary phase shift between Earth terms and the pulsar term for a specified pulsar. Our algorithm automatically searches for that phase shift and thus corrects the localization bias caused by pulsar terms in an Earth-term search.
\item our results indicate only a small increase in the detection probability compared with a method that does not account for the pulsar term.
\item the method is fast and can be applied to large data sets.
\item the code implementing this algorithm is available at the \href{http://www.atnf.csiro.au/research/pulsar/ppta/}{PPTA Wiki page}.
\end{itemize}

The present method is designed for non-evolving binaries in circular orbits. SMBHBs detectable in the PTA band may (a) experience non-negligible frequency evolution over Earth-pulsar light-travel times and (b) have significant eccentricities. In the former case, a successful search can provide estimation of the binary chirp mass \citep{Taylor14}. Data analysis techniques for eccentric binaries have been developed in \citet{Zhu15Single} and \citet{Taylor16ecc}, both of which consider only the Earth terms. How the sky localization is biased in such searches and how to correct for/remove such a bias should be addressed in future studies.

\section*{Acknowledgments}
X-JZ and LW acknowledge funding support from the Australian Research Council and computing support from the Pawsey Supercomputing
Centre at WA. GH is supported by an Australian Research Council Future Fellowship. YW acknowledges support by the National Science Foundation of China (NSFC) through award number 11503007. The contribution of SDM is supported by NSF grants PHY-1505861 and HRD-1242090.

\bibliographystyle{mn2e}
\bibliography{Ref}
\label{lastpage}
\end{document}